# Time-Domain Studies of Very-Large-Angle Magnetization Dynamics Excited by Spin Transfer Torques

I. N. Krivorotov[1], N. C. Emley[2], R. A. Buhrman[2], D. C. Ralph[2]

[1]Department of Physics and Astronomy, University of California, Irvine, California 92697-4575, USA.

[2]Cornell University, Ithaca, New York 14853-2501, USA.

We describe time-domain measurements which provide new information about the large-angle nonlinear dynamics of nanomagnets excited by spin-transfer torque from a spin-polarized current. Sampling-oscilloscope measurements, which average over thousands of experimental time traces, show that the mean reversal time for spin-transfer-driven magnetic switching has a step-like dependence on magnetic field, because an integer number of precession cycles is required for reversal. Storage-oscilloscope measurements of individual experimental traces reveal non-periodic large-amplitude resistance variations at values of magnetic field and current in a crossover region between the regimes of spin-transfer-driven switching and steady-state precession. We also observe directly the existence of time-dependent switching, on the nanosecond scale, between different precessional modes and between a precessional mode and a static state, at particular values of magnetic field and current bias.

PACS:    72.25.Ba, 75.75.+a, 75.20.-g

# I. Introduction

The spin-transfer torque from a spin-polarized direct current interacting with a nanomagnet can produce several different types of magnetic dynamics,[1-10] including switching between static magnetic states[11-30] and the generation of steady-state precession.[31-51] In order to understand the nature of the modes that can be excited, the phase diagram of spin-transfer-driven dynamics has been mapped as a function of current and the angle and magnitude of an applied magnetic field.[52-62] The primary experimental techniques applied in this effort have been frequency-domain measurements of the spectra of resistance oscillations excited by a direct current.[54,55,57,58,61] While very important, frequency-domain measurements are primarily useful for studying signals that are approximately periodic or that exhibit random telegraph-like switching dynamics.[63] They can provide little insight into transient dynamics or other non-periodic signals. In this Article, we report the results of new experimental approaches that employ direct time-domain electrical measurements to probe the transient dynamics active in spin-transfer-driven magnetic switching and to explore for the existence of persistent but non-periodic current-driven magnetic states. This work builds on a previous publication by our group.[64] Here we use both sampling-oscilloscope measurements that average over thousands of experimental traces and storage-oscilloscope measurements of single traces, in an effort to characterize the full phase diagram of spin-transfer-driven dynamics as a function of current, magnetic field, and the time following a current step. We are able to demonstrate clearly that spin-transfer-driven magnetic switching occurs as the result of process in which the moment of a nanomagnet precesses to larger and larger angles before reversing. This results in an average reversal time that shows a step-like dependence on the magnitude of applied magnetic field because an integer number of precessional cycles is required for reversal, supporting a previous conclusion by Devolder *et al.*[65] Our measurements also provide a direct view of several different types of persistent but non-periodic spin-transfer-driven magnetization dynamics, including aperiodic, large-angle magnetic rotations for values of current and magnetic field in a crossover region between the regimes of switching and steady-state precession, and fast, nanosecond-scale switching between different precessional modes and between a precessional mode and a static magnetic state.

## II. Device Fabrication and Characterization

We employ devices of the same design used in Ref. [64] and [66]. Sample fabrication begins with high-vacuum magnetron sputtering of a magnetic multilayer consisting of Cu (80 nm) / $Ir_{20}Mn_{80}$ (8 nm) / Py (4 nm) / Cu (8 nm) / Py (4 nm) / Cu (20 nm) / Pt (30 nm) onto an oxidized Si wafer (Py = permalloy = $Ni_{80}Fe_{20}$). The deposition is done at room temperature with a 500 Oe magnetic field applied in the plane of the sample, and the multilayer is annealed at $T$ = 250 °C for 80 minutes in the same field. A subtractive nanofabrication process is then used to define spin valves of approximately elliptical cross section with major and minor diameters of 130 nm and 60 nm, and with Cu electrodes making contact to the top and bottom of the structure. The nominal direction of the exchange bias field from the $Ir_{20}Mn_{80}$ layer, set during the deposition and subsequent annealing, is in the plane of the sample at 45° with respect to the major axis of the ellipse. We will call the Py layer that is coupled to the exchange-bias field of the $Ir_{20}Mn_{80}$ the "pinned layer" and the other Py layer the "free layer". The purpose of the exchange bias is to set a controlled, non-zero offset angle $\theta$ between the orientation of the pinned and free-layer magnetizations, because the spin-transfer torque goes to zero for $\theta = 0$.[1] All data in this paper except that in Fig. 4(d) were obtained from one most-studied sample, although similar behavior was observed for other samples from a set of forty.

We performed all measurements reported in this paper using an initial sample temperature of 4.2 K, although Ohmic heating resulted in the actual sample temperature rising to 30 K – 60 K upon application of currents in the range 5-10 mA.[66, 67] We applied a magnetic field, $H$, in the plane of the sample at 45° with respect to the ellipse major axis and 90° from the exchange-bias direction (see Fig. 1(a)). The hysteresis curve of differential resistance as a function of magnetic field, exhibiting the field-driven switching characteristics of the free layer, is shown in Fig. 1(b). This field dependence can be fit successfully to macrospin Stoner-Wohlfarth simulations of the magnetic orientation, from which we determine that the difference in zero-bias resistance between the fully parallel and fully antiparallel magnetic configurations is $\Delta R_{GMR}$ = 0.16 Ω.[66] Because of the exchange bias, the moments of the two magnetic layers at zero applied field are not fully parallel or antiparallel, so we will refer to the two stable magnetic configurations as simply the low-resistance (LR) and high-resistance (HR) states. As shown in Fig. 1(c), the spin-transfer torque from a direct current $I$ flowing through the sample at $|H| \leq 330$ Oe can drive hysteretic switching of the free layer magnetization between the LR and HR states, with critical currents at $T$ = 4.2 K and $H$ = 0 of $I_c^+$ = 2.3 mA and $I_c^-$ = -1.5 mA. (Note that the sample geometry here is not designed to minimize the switching currents, in contrast to other recent work.[24,25]) Positive current corresponds to electron flow from the free to the pinned layer. For $H$ > 330 Oe and $I \geq 2.7$ mA, the low-frequency differential resistance exhibits a series of peaks (Fig. 1(c)). These peaks correspond to transitions between different modes of persistent oscillation for the magnetization of the free layer.[54,66] The DC resistance of the sample in this regime of magnetization dynamics increases monotonically from the LR value to the HR value as a function of $I$ (Fig. 1(d)).

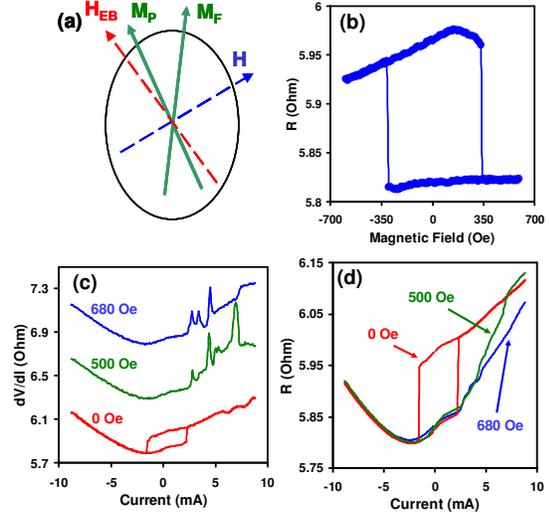

Figure 1. (Color online) (a) Schematic top view of the spin valve showing the directions of the magnetizations of the pinned, $M_P$, and the free, $M_F$, layers for a positive applied magnetic field **H**. $H_{EB}$ is the exchange-bias field acting on the pinned layer. (b) Resistance of the spin valve at zero bias as a function of external magnetic field at 4.2 K. (c) Differential resistance of the spin valve as a function of direct current at 4.2 K measured at different values of $H$. The curves for $H$ = 500 Oe and 680 Oe are offset vertically. (d) DC resistance of the sample as a function of bias current obtained from the data in (c) by numerical integration.

### III. Sampling-Oscilloscope Measurements of Switching and Precession

We first report measurements of current-driven magnetization dynamics made using a 20-GHz sampling oscilloscope, with a technique similar to that employed in Ref.

[68] and [64]. We initialize the sample in its LR state by a negative current pulse. We then apply a positive current step with a 150-ps rise time to excite magnetization dynamics. The voltage across the sample is amplified with a 30-dB 15-GHz amplifier, recorded by the sampling oscilloscope triggered by the same source used to drive the current step, and averaged over 20,000 oscilloscope traces. After a background subtraction procedure described in Ref. [64] and corrections for signal attenuation and amplification in the microwave circuit, we plot the voltage signal corresponding to:

$$V(t) = I(t)(R(t) - R_{HR})\frac{50\ \Omega}{2R_S + R_L + 50\ \Omega}. \quad (1)$$

Here $I(t)$ is current through the spin valve, $R(t)$ is time-dependent resistance of the full sample, $R_{HR}$ is resistance of the full sample in the HR state, $R_L$ (26 Ω) is resistance of the leads and $R_S$ (5.9 Ω) is the average resistance of the nanopillar excluding leads.[64] Examples of the voltage signal obtained for a current step magnitude of 8.2 mA and two representative values of magnetic field are shown in Fig. 2(a). The initial decrease of these signals is due to the 150-ps rising edge of the current step, $I(t)$, and is not related to a change of the magnetic state of the sample. Once the current step is fully applied, the initial negative value of the voltage in Fig.

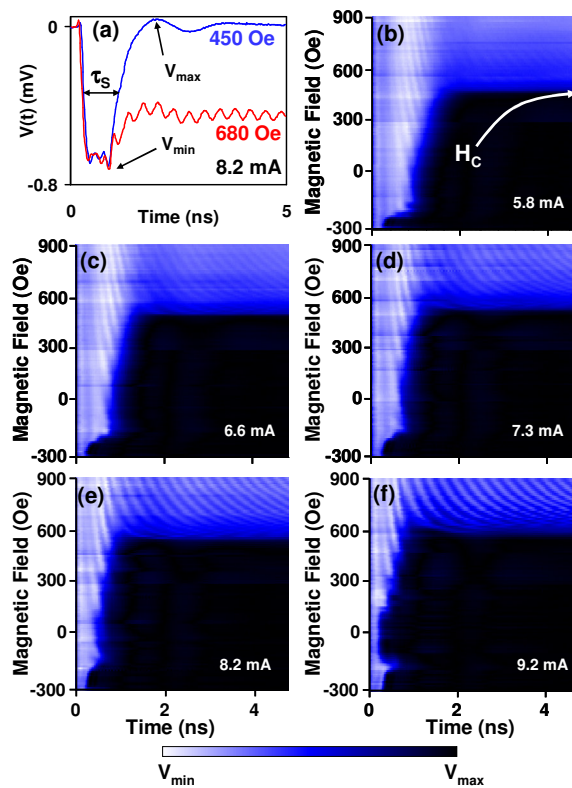

Figure 2. (Color online) (a) Average voltage signals at 4.2 K due to current-induced dynamics of the free-layer nanomagnet as measured by a sampling oscilloscope, in response to a 8.2-mA current step at two different values of the external magnetic field: $H$ = 450 Oe (current-induced switching) and $H$ = 680 Oe (current-induced persistent precession). Each plot shows a signal averaged over 2×10[4] traces. The definition of the switching time, $\tau_S$, is illustrated for the $H$ = 450 Oe signal. (b-f) Magnetic-field dependence of average voltage traces due to current induced dynamics for five different amplitudes of current step, as marked. Color scale: white corresponds to the minimum voltage, $V_{min}$, black corresponds to the maximum voltage, $V_{max}$ of the signal. ($V_{max}$ and $V_{min}$ differ from panel to panel, due to the changing current bias.) The zero of the time scale is set to be at the midpoint of the leading edge of current step, ($V_{max}$-$V_{min}$)/2.

2(a) corresponds to the LR state of the spin valve. (As noted in Eq. (1), the HR state is used as the reference value for the resistance.) The oscillations of the voltage signal evident in Fig. 2(a) indicate that the current step causes the magnetization of the free layer to precess, changing the sample resistance. For $H = 450$ Oe, the free layer switches to the HR state after a few oscillations, causing $V(t)$ to go to zero. However, for $H = 680$ Oe, the oscillations become persistent and switching to the HR state does not take place.

Panels (b)-(f) in Fig. 2 show compilations of such voltage signals due to current-driven magnetization dynamics as a function of magnetic field, for several different current-step magnitudes. The initial response is qualitatively similar for all of the magnetic field magnitudes measured -- we observe resistance oscillations whose amplitude and frequency vary smoothly as a function of $H$. However, after approximately 0.3-1 ns, the dynamics eventually separate into two classes, with switching to the HR state occurring for magnetic fields less than a critical field $H_c(I)$, and with persistent precession for $H > H_c(I)$. This bifurcation is consistent with previous phase diagrams of the magnetization dynamics, which show separate regimes of switching and persistent precession.[52,54] The precessional nature of the dynamics prior to switching has important consequences for the mean reversal time, $\tau_s$. We define $\tau_s$ as the time interval between the midpoint of the rising edge of the current step and the midpoint of the transition between the LR and HR states (see Fig. 2(a)), and we plot its value as a function of $H$ for $I = 9.2$ mA in Fig. 3(a). We find that the mean reversal time increases in a discrete, stepwise manner as a function of $H$ for $0 < H < H_c$. Each step coincides with an extra voltage oscillation in Fig. 2. This clearly demonstrates that the reversal process proceeds via precession of the magnetization with increasing amplitude, and an integer number of precession cycles of the free layer magnetization is required for the reversal to take place. This result agrees with conclusions of Devolder *et al.* who observed stepwise changes in switching probability as a function of the length of short current pulses.[65] The number of precession cycles required before reversal is greater for large $H$ because the field increases the energy barrier for the LR→HR transition, and thus more time is needed for spin torque to transfer enough energy to surmount the barrier.

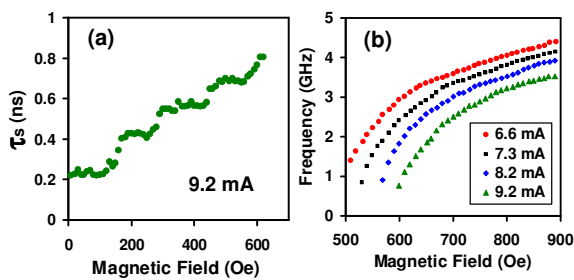

Figure 3. (Color online) (a) Reversal time $\tau_S$ as a function of magnetic field for $I = 9.2$ mA. (b) Frequency of the oscillatory signal in the regime of persistent dynamics as a function of $H$ obtained from the sampling-oscilloscope measurements for four amplitudes of the current step.

Another notable feature of the data in Fig. 2(c)-(f) is that the steady-state precession frequency approaches zero for $H \to H_c$ from above. This is evident from the way that the signal minima shift to the right and their spacing increases as $H$ is reduced from large values. The field-dependence of the precession frequencies extracted from these measurements is shown quantitatively in Fig. 3(b). The vanishing frequency at the transition between the regimes of switching and persistent precession is consistent with the conventional spin-torque picture of current-driven dynamics. For $H < H_c$, the magnetic energy of the free-layer magnet, $E(\bm{m}_f)$, as a function of the direction of its magnetization, $\bm{m}_f$, has two local minima corresponding to the LR and HR states, while for $H > H_c$ it has a single minimum because the HR state is no longer stable in the absence of applied current. At $H = H_c$, the curvature of the magnetic energy surface $E(\bm{m}_f)$ must go to zero in the neighborhood of the disappearing minimum, and thus the effective magnetic field $\bm{H}_{eff} = -\nabla E(\bm{m}_f)$ and the frequency of magnetization precession should also be small, *as long as the precession angle is sufficiently large to approach the disappearing local minimum* near the magnetization angle corresponding to the HR state. From the macrospin Stoner-Wohlfarth simulations, we estimate that the angle between the global energy minimum and the disappearing local minimum varies between 175° and 124° as the external field varies between 0 Oe and 320 Oe,[66] so that when the magnetization of the free layer starts in the LR state, the magnetic trajectory must be repeatedly driven to a very large angle when $H$ is near $H_c$. Additional information about the amplitude of magnetic precession can be obtained by analyzing the size of the oscillatory voltage signal, but we will defer this discussion until section V, where we compare the amplitudes of oscillation measured by several different time-domain techniques.

## IV. Storage-Oscilloscope Measurements of Large-Angle Dynamics

The sampling-oscilloscope measurements described in the previous section necessarily average over many repeated traces due to the nature of the sampling measurement technique and to improve the signal-to-noise ratio. This provides a useful view of transient dynamics that are reproducible, but any differences between traces average away and are lost from the signal. In this section, we describe the use of a 20-Gigasample/second storage oscilloscope to study individual traces of the voltage signals due to spin-transfer-driven magnetic dynamics. We observe that there are regions of the dynamical phase diagram where the magnetization does not move reproducibly, but instead can exhibit aperiodic pulses and random fluctuations between different dynamical modes.

In our storage-oscilloscope measurements, we simply apply a direct current to the sample (not pulsed), amplify the voltage across the sample with a 30-dB 15-GHz amplifier, and record the signal in real time with the storage oscilloscope. Results for the same sample studied in the previous section are shown in Fig. 4(a)-(c) for $H$ = 500 Oe and three different current biases (5.8 mA, 6.6 mA and 7.0 mA). In Fig. 4, we plot

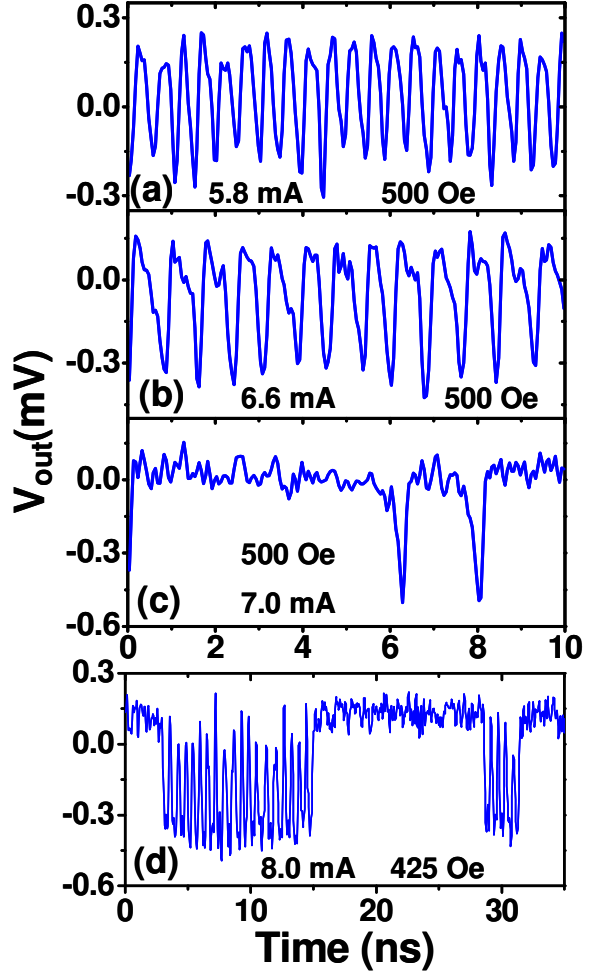

Figure 4. (Color online) Real-time signal due to magnetization dynamics measured with a microwave storage oscilloscope at $H$ = 500 Oe for three values of the current bias (a) $I$ = 6.0 mA, (b) $I$ = 6.8 mA, and (c) $I$ = 7.0 mA. (d) Real-time signal from a different sample showing telegraph-type switching between a static high-resistance state and a mode with large-amplitude persistent precession.

$$V_{out}(t) = R(t)I\frac{50\ \Omega}{2R_S + R_L + 50\ \Omega} - constant. \qquad (2)$$

The root-mean-square background noise (due to Johnson noise and amplifier noise) for these broadband measurements is ~ 0.1 mV, which is a significant fraction of the total voltage signals shown in Fig. 4. However, despite this relatively large noise, the signals generated by motion of the free-layer magnetization can still be observed when $I$ is sufficiently large. At $I = 5.8$ mA (Fig. 4(a)), the dynamics of the free layer appear to be periodic and the resistance variations are approximately sinusoidal. This is as expected from the sampling oscilloscope measurements in Fig. 2. However, at larger current values, for which the bias point comes closer to the boundary $H_c(I)$ between the precessional and switching regimes, the dynamics acquire a degree of non-periodicity and become non-sinusoidal. (The average frequency of oscillations also decreases, in agreement with the sampling oscilloscope measurements.) For example, at $I = 6.6$ mA, the sample spends significantly more time close to the highest voltage value than to the lowest voltage value, with relatively fast swings to low voltage and back. This motion is still approximately periodic, but small random variations of the period are visible in Fig. 4(b). At $I = 7.0$ mA, very close to the boundary between precession and switching, the dynamics become strikingly non-periodic (Fig. 4(c)). For the great majority of time, the voltage signal is approximately constant at a value that corresponds to the HR state of the sample. At apparently random intervals, the voltage departs from this value, and the sample resistance makes brief excursions, producing a voltage change that corresponds to approximately 75 % of the full resistance difference $\Delta R_{GMR}$ determined at zero bias. Such a large variation of resistance gives a lower bound of 120° on the peak-to-peak amplitude of in-plane angular excursions of magnetization of the free layer, which is close to the estimated angular separation of the two minima of $E(m_f)$ for $H < H_c$. Despite the randomness in timing, the dynamics still have some regularity, in the sense that the shape of the time-dependent voltage waveform during the excursions is similar for each excursion.

We believe that this non-periodic motion can be explained by the passage of magnetization close to the vanishing minimum of magnetic energy $E(m_f)$, where magnetic dynamics become slow. As noted above, this corresponds to an orientation of the free-layer magnetization near a weakly-unstable HR state. The non-periodic dynamics can result from an enhanced effect of thermal fluctuations in the vicinity of the flat region of $E(\boldsymbol{m}_\text{f})$. Here the random Langevin field due to thermal fluctuations becomes comparable to the deterministic effective field $\boldsymbol{H}_\text{eff} = -\nabla E(\boldsymbol{m}_\text{f})$. Under these conditions, the motion of the magnetization may take the form of a slow random walk in the neighborhood where $\boldsymbol{H}_\text{eff} \approx 0$. Only when the magnetization direction diffuses sufficiently far away from the region of small $\boldsymbol{H}_\text{eff}$, after a random time interval spent in this neighborhood, will the torque due to $\boldsymbol{H}_\text{eff}$ drive the motion along a deterministic trajectory along a large angle orbit. After one cycle, the free layer moment returns to the region where $\boldsymbol{H}_\text{eff} \approx 0$, and the process begins again.

We do wish to note that the behavior shown in Fig. 4(c) is not universal for all samples biased near $H_c(I)$. Fig. 4(d) shows an example of the magnetization dynamics observed in another nominally identical sample for similar bias conditions. For this sample, the magnetization undergoes two-state switching at random time intervals between the static HR state and a mode of large-angle persistent precession. The fact that qualitatively different types of dynamics can be observed for similar samples suggests that the magnetic dynamics under these bias conditions are very sensitive to the details of the magnetic energy landscape, $E(\boldsymbol{m}_\text{f})$. In previous work our group has suggested the possibility of random telegraph switching between static and dynamic magnetic states near boundaries of the dynamical phase diagram.[67] The data in Fig. 4(d) directly demonstrate this type of switching on the few-ns time scale.

## V. Storage-Oscilloscope Measurements Triggered by the Signal Itself

A more precise measurement of the waveform for the voltage excursions shown in Fig. 4(c) would be useful for future tests of micromagnetic simulations for the very-large-angle dynamics of samples biased near $H_c(I)$. However, the background noise in our single-trace storage oscilloscope measurements makes it difficult to make more precise measurements by this technique. Likewise, standard sampling oscilloscope measurements, using a trigger provided by the same current pulser that initiates the magnetic dynamics, also become increasingly inaccurate for biases near $H_c(I)$ because the non-periodic nature of the dynamics cause the oscillatory part of the signal to disappear when averaging over many repeated traces. To overcome these difficulties, we have measured the magnetic dynamics using a storage oscilloscope triggered not by the current pulse that initiates the dynamics, but by an amplified copy of the signal itself. This enables reliable averaging of transient waveforms that are reproducible but not necessarily periodic, with a time resolution limited by the noise-induced jitter error of our triggering, approximately 10 ps for a signal with the peak-to-peak amplitude of 0.5 V.

As a test of the method, Figure 5(a) shows an example of this type of measurement for the bias conditions $H = 680$ Oe and $I = 6.6$ mA, sufficiently far away from the phase-boundary region $H_c(I)$ that the persistent dynamics should be approximately periodic. The oscillations are indeed resolved much better and exhibit a larger amplitude compared to the standard sampling-oscilloscope measurement shown in Fig. 2(a) and Fig. 2(c). Figure 5(b) shows the averaged result for the first period of oscillation for $H = 500$ Oe and $I = 5.8$ mA. For bias conditions such as this, well away from $H_c(I)$, the resistance signal is approximately sinusoidal.

Figure 5(c) displays the averaged voltage signal for the interesting case of large angle excursions in the phase-boundary region near $H_c(I)$, where Fig. 4(c) exhibits strikingly non-periodic signals. In this case, the resistance waveform is strongly non-sinusoidal. Starting from the HR state near the beginning of the trace in Fig. 5(c), the signal decreases relatively slowly to a minimum, and then returns much more abruptly back to the HR state, with a total excursion time of approximately 1.7 ns. The same asymmetry in the shape of the waveform is visible, albeit less clearly, in Figures 4(b) and (c). The overall amplitude of the voltage change corresponds to approximately 65% of the full resistance difference $\Delta R_{GMR}$ at zero bias.

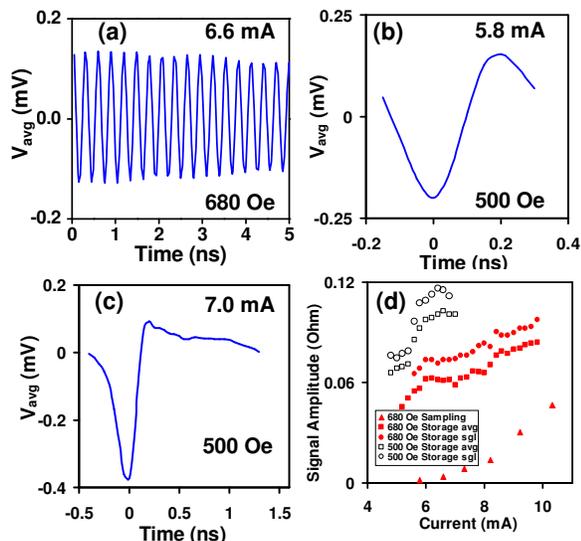

Figure 5. (Color online) (a) Average signal (over $2\times10^4$ traces) due to persistent magnetization dynamics measured by a storage oscilloscope triggered by the signal itself, for $H = 680$ Oe and $I = 6.6$ mA. (b) The first period of the averaged oscillatory signal at $H = 500$ Oe and $I = 5.8$ mA. (c) The averaged signal in the regime of non-periodic dynamics, for $H = 500$ Oe and $I = 7.0$ mA. (d) Comparison of the current dependence of the oscillatory signal peak-to-peak amplitudes measured by the three different time-domain techniques described in the text: (i) sampling oscilloscope measurements of dynamics in response to a current step (triangles), (ii) storage oscilloscope measurements employing the signal itself for triggering (squares), and (iii) single-trace storage oscilloscope measurements (circles). Solid symbols are data for $H = 680$ Oe, open symbols are data for $H = 500$ Oe. The full difference in resistance between parallel and antiparallel magnetic configurations for the sample is $\Delta R_{GMR} = 0.16$ $\Omega$ at low bias.

A comparison of the amplitudes of the oscillatory signals obtained by the three different time-domain techniques described in this article is shown in Fig. 5(d), for two different values of applied magnetic field. In this figure, the measured voltage signals are converted to the corresponding resistance oscillations, so that they can be understood relative to the full difference in resistance between the parallel and antiparallel magnetic configurations, $\Delta R_{GMR} = 0.16$ Ω. Of the three measurement techniques, the oscillation amplitude is largest in the single-trace storage-oscilloscope measurements, because this measurement is least susceptible to thermal fluctuations and timing errors. The mean amplitude of the resistance oscillations measured by single-trace measurements is as large as 75% of $\Delta R_{GMR}$. As noted above, this corresponds to a very large precession angle, at least 120°. Next largest are the signals measured by the storage oscilloscope triggering on the signal itself. These signals are typically 15-20% smaller than those obtained from the single-trace measurements. The reduced average signal amplitude is likely due primarily to jitter error in the oscilloscope triggering. The amplitude of the oscillatory signals measured by the sampling-oscilloscope technique (triggering on the pulser that provides the current step used to excite the magnetic dynamics) can be significantly smaller than measured by the other two methods. This is because the sampling-oscilloscope measurements are averages over many experimental traces, each of which can have a slightly different phase of precession due to thermal fluctuations in the initial magnetization angle and during the initial stage of small-amplitude magnetization precession following application of the current step. No oscillatory signal can be observed by the sampling-oscilloscope technique for $I < 5.8$ mA in Fig. 5(d), even though frequency-domain measurements with a spectrum analyzer show clearly that steady-state precessional dynamics exist for $I > 2.7$ mA. The differences between the oscillation amplitudes measured by the three techniques decreases as a function of increasing current, so that by $I = 10$ mA, the signals differ only by about 45%. We believe that the effects of dephasing are reduced for large currents in the sampling-oscilloscope measurements because large currents result in a faster transition to the large-amplitude regime of dynamics following the current step. As a result, thermal fluctuations in the initial stage of magnetization precession lead to smaller phase fluctuations in the resistance signal. This picture of reduced dephasing at large current amplitude is supported by theoretical calculations that predict an inverse proportionality between the dephasing rate and square of the amplitude of precession.[69, 70]

## VI. Measurements of Time-Dependent Switching Between Different Precessional Modes

In the sections above, we have described the existence of non-periodic magnetic dynamics near one boundary in the dynamical phase diagram for spin-transfer-driven motion -- at values of $I$ and $H$ near the critical field $H_c(I)$ between static switching and persistent precession. Within the regime of persistent precession, there are additional boundaries at which the dominant precessional mode changes, producing jumps in frequency.[54,55,58,66,69] Figure 6(a) shows the oscillation frequency as a function of current at three values of magnetic field for our most-studied sample, as determined by frequency-domain spectrum-analyzer measurements.[66] At $H = 680$ Oe, as a function of increasing $I$, the measured frequency undergoes two jumps downward, near 3.7 mA and 4.8 mA. As $H$ is decreased to 600 Oe and 500 Oe, the current values where the jumps occur also decreases. We have argued previously, based on micromagnetic simulations, that these jumps are associated with transitions between different non-linear magnetic modes.[66,71] At bias points near each transition, both the higher and lower frequencies can be seen together in the dc-driven oscillation spectrum when it is integrated over a time scale of seconds to minutes (Fig. 6(d)). In this section, we discuss the use of time-domain techniques to explore in more detail the ns-scale magnetic dynamics near these transitions.

When we attempt to use a sampling or a storage oscilloscope to measure the signals near these bias points, using either standard triggering from the current step or triggering from the signal itself, the result after averaging over many traces is similar to that shown in Fig. 6(b). The averaged oscillations exhibit a beating pattern, showing the existence of two different characteristic frequencies. However, from this type of averaged signal, it is impossible to determine if the two modes with different frequencies are excited simultaneously or if only one mode or the other is excited at a given moment of time.

To resolve the nature of the dynamics more clearly, we can employ single-shot storage-oscilloscope measurements. Fig. 6(c) shows a typical single trace spanning 20 ns, for the bias conditions ($I$ = 4.5 mA and $H$ = 600 Oe) at which a frequency jump is observed in the spectrum shown in Fig. 6(a). Because the mode transition takes place at a relatively low value of current bias, the electronics noise in the single-shot measurement is comparable to the amplitude of the oscillatory signal, so that direct inspection of Fig. 6(c) does not allow us to determine if the two modes coexist or are excited sequentially. However, we can distinguish between these two possibilities using Fourier analysis of the data. Figure 6(e) shows the fast Fourier transform (FFT) of the full 20-ns long time trace in Fig. 6(c). This spectrum exhibits both of the oscillation frequencies present in the 2-minute spectrum in Fig. 6(d). To search for the presence of time periods in which just a single mode might be excited, we performed FFT analysis of the signal over shorter time intervals (1.5 – 5 ns) for different parts of the 20-ns trace. Figures 6(f) and (g) show FFTs of different 4-ns long segments of the data in Fig. 6(a). Only the lower-frequency mode is excited in the interval from 0 ns to 4 ns and while only the higher frequency mode is excited from 16 ns to 20 ns. We have analyzed many time intervals of variable duration for many different 20-ns signal traces. The data invariably show the existence of time intervals where either one or the other mode is excited, along with intervals where the FFT shows the presence of both modes. We found no intervals in which neither mode was present. From these measurements we conclude that the magnetic dynamics near the frequency jumps shown in Fig. 6(a) consist of switching on the ns time scale in which different precessional modes alternate between being active and inactive.

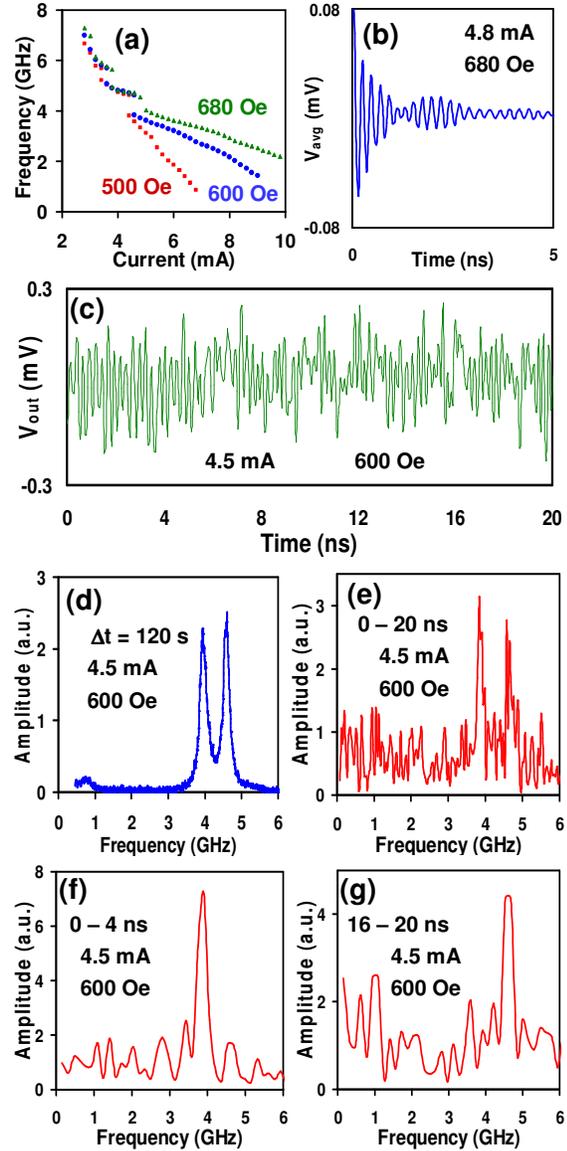

Figure 6. (Color online) (a) Frequencies of dc-driven magnetic dynamics measured with a spectrum analyzer, for three values of the external magnetic field, showing discontinuities in frequency as a function of current within the regime of persistent magnetic dynamics. (b) Average signal (over $2\times10^4$ traces) due to persistent magnetization dynamics measured by a storage oscilloscope triggered by the signal itself, for $H$ = 680 Oe and $I$ = 4.8 mA, near one of the frequency discontinuities. (c) Real-time signal captured by a microwave storage oscilloscope for $H$ = 600 Oe and $I$ = 4.5 mA. (d) Amplitude spectrum of the voltage signal due to magnetic dynamics, measured with a spectrum analyzer with an averaging time of two minutes, for the same bias conditions as in (c). (e) Fast Fourier transform of the full 0-20 ns signal in (c). (f) Fast Fourier transform of the data in (c) from 0 to 4 ns. (g) Fast Fourier transform of the data in (c) from 16 to 20 ns.

In previous work,[69] we proposed that random telegraph noise between two dynamic states may be a significant (sometimes the dominant) mechanism producing dephasing and therefore limiting the linewidth of the dc-current-driven oscillations, since the phase of oscillations is not maintained in the random transitions between the modes. The data in Fig. 6 directly prove that such mechanism is at play in our spin transfer devices. Indeed, the linewidth of the spectral peaks in Fig. 6(d,e) significantly exceeds the typical linewidth at currents far from the mode transition regions.[66]

## VII. Conclusions

We have described the results of several different time-domain measurement techniques which provide new understanding of the large-angle magnetization dynamics of a nanomagnet driven by spin transfer torque that cannot be obtained by conventional frequency-domain spectroscopy measurements.

In the current-induced switching regime, we observe a step-wise dependence of the magnetization reversal time on the magnitude of an external magnetic field. This step-wise dependence arises from the requirement to have an integer number of precession cycles in order to reach the bifurcation point for reversal.

Near the phase boundaries of the current-field phase diagram that separate different static and dynamic states of magnetization, our measurements reveal that the magnetization dynamics can become stochastic. In particular, close to the phase boundary between persistent oscillatory dynamics and current-driven switching, we find that the resistance signals due to the persistent dynamics evolve as a function of increasing current at a fixed applied field, from a periodic sinusoidal oscillation to a signal consisting of brief, randomly-timed, and distinctly non-sinusoidal swings in resistance, with an amplitude as large as 75% of $\Delta R_{GMR}$. Simultaneously, the average frequency of the oscillations approaches zero. These measurements suggest that the magnetization dynamics are affected by thermal fluctuations in the neighborhood of the shallow local minimum of the magnetic energy landscape that exists near the static-dynamic phase boundary for a very large precession angle.

The magnetization motion at boundaries of the dynamical phase diagram separating different modes of persistent oscillation can also have a stochastic character, and can be described as random switching between two dynamics modes having different frequencies. This switching can occur on a time scale of nanoseconds and it is the dominant mechanism limiting the linewidths of the dc-driven persistent oscillations near these phase boundaries.

The authors thank J. Sankey, J. Miltat, D. Mills, A. Slavin and Y. Tserkovnyak for many useful discussions. This research was supported by the Office of Naval Research, and the National Science Foundation's Nanoscale Science and Engineering Centers program through the Cornell Center for Nanoscale Systems. We also acknowledge NSF support through use of the Cornell Nanoscale Facility node of the National Nanofabrication Infrastructure Network and the use of the facilities of the Cornell Center for Materials Research.